\documentclass[superscriptaddress, nofootinbib,reprint]{revtex4-1}
\usepackage{amssymb}
\usepackage{amsmath}
\usepackage{times}
\usepackage{graphicx}
\usepackage{epsfig}
\usepackage{fancyhdr}
\usepackage{float}
\usepackage[colorlinks,citecolor=blue, urlcolor=blue, linkcolor=blue]{hyperref}

\def\turbo{{TURBO\ }}
\def\uu{\mathbf u}
\def\bb{\mathbf b}
\def\zz{\mathbf z}
\def\ff{\mathbf f}
\def\xx{\mathbf x}
\def\kk{\mathbf k}

\begin{document}

\title{Controlling the level of the ideal invariant fluxes for MHD turbulence using \turbo spectral solver}
\author{B.~Teaca}
\email{bogdan.teaca@epfl.ch, Tel: +41-21-693.43.05}
\affiliation{Centre de Recherches en Physique des Plasmas, Science de Base, Ecole Polytechnique Federale de Lausanne, Station 13, Building PPB
CH-1015 Lausanne, Switzerland.}
\author{C.~C.~Lalescu}
\email{clalesc1@jhu.edu}
\affiliation{Department of Applied Mathematics \& Statistics, The Johns Hopkins University, 3400 North Charles Street, Baltimore, Maryland 21218-2682 USA.}
\author{B. Knaepen}
\email{bknaepen@ulb.ac.be}
\affiliation{Statistical and Plasma Physics, Faculty of Sciences, Universit\'e Libre de Bruxelles, Campus Plaine, CP 231, B-1050 Brussels, Belgium.}
\author{D.~Carati}
\email{dcarati@ulb.ac.be}
\affiliation{Statistical and Plasma Physics, Faculty of Sciences, Universit\'e Libre de Bruxelles, Campus Plaine, CP 231, B-1050 Brussels, Belgium.}
\begin{abstract}
The ideal invariants present in the formalism of magnetohydrodynamics (MHD), i.e. global quantities that are conserved in the absence of sources and dissipative effects, play an important role in various theoretical and numerical studies of MHD turbulence. The fluxes of these ideal invariants represent separate channels that transfer the information across different scales in a turbulent system. Once a statistically stationary state of turbulence is reached, the amount of any ideal invariant quantity introduced in the system by a forcing mechanism equals the amount of the same quantity removed by the dissipative effects from the system. For highly developed turbulence, these two mechanisms act predominantly at different scales that are largely separated. Since the ideal invariant quantities cascade between scales, a constant flux is generated with great implication on the state of the system. Numerically, controlling the ideal invariant fluxes levels for a turbulent MHD system is important for the analysis of fundamental MHD turbulence properties. We propose a forcing mechanism that controls the three ideal invariants of MHD turbulence: the total energy, the cross-helicity and the magnetic helicity. This forcing is implemented in the freely available \turbo solver, that is also briefly presented. 
\end{abstract}
\maketitle
\onecolumngrid

%%%%%%%%%%%%%%%%%%%%%%%%%%%%%%%%%%%%%%%%%%%%%%%%%%%%%%%%%%%%%%%%%%%%%%%%%%%%%%%%
%%%%%%%%%%%%%%%%%%%%%%%%%%%%%%%%%%%%%%%%%%%%%%%%%%%%%%%%%%%%%%%%%%%%%%%%%%%%%%%%
\section{Introduction}

The motion of a fluid is described mathematically by a nonlinear evolution equation, for which no general analytical solution exists. Except for laminar flows in very simple geometries, even the use of a perturbative approach is impossible. This known problem, caused by the coupling of different scales in a flow by the nonlinear terms has triggered the use of numerical methods in the study of fluid flows. Usually, solutions of a flow are obtained using a numerical solver, in an incremental manner, starting from a set of initial conditions and a prescribed set of boundary conditions. For a typical engineering problem, the correct description of the physical properties and geometry of the boundary conditions is crucial, as they are responsible for an entire class of instabilities introduced in the flow which in turn characterizes the subsequent motion. However, for the study of fully developed turbulence, where a huge range of scales are present and a clear separation exists between the geometry dependent large scales and the universal, dissipative small scales, periodic boundary conditions represent the preferred choice. In this situation, the driving instabilities are introduced in the form of an external force that acts only at a particular scale, usually a large one. The use of periodic boundaries conditions for a flow has the advantage of enabling us to easily translate the motion problem in terms of Fourier modes which can be solved numerically using spectral solvers. For this purpose we make use of the \turbo code, which we will briefly introduce in Section \ref{sec2} of this paper.

When the fluid is electrically conducting like in the case of a plasma or of a liquid metal, the momentum balance equation is influenced by the Lorentz force and the number of non-linear terms increases compared to the case of fluid turbulence. This coupling of the Navier-Stokes equation with the Maxwell reduced equations, which leads to the magnetohydrodynamic (MHD) equations, increases the complexity of the interaction between scales. Compared to hydrodynamical flows, the ideal invariants change in the case of MHD, as the interplay between the velocity and the magnetic fields needs to be considered. The importance of the MHD ideal invariants will be discussed in Section \ref{sec3}. 

MHD turbulence, considered as an initial value problem with periodic boundary conditions, tends to decay in absence of driving mechanism. To maintain a stationary turbulent state, it is needed to add sources in the evolution equations that mimic the effects of the various instabilities that may appear in the large, geometry dependent scales of a realistic system. One way of achieving this would be through a constrain on the large scale value of one or both fields. This is often implemented in spectral space by freezing the value of a limited set of Fourier modes characterizing the largest scales in the system~\cite{Muller:2007p755, Mason:2008p69}. Alternatively, an external artificial force can be used \cite{Alexakis:2005p304, Teaca:2009p628}. We propose such a force in Section~\ref{sec4} which controls the injection rate of the three ideal invariants of MHD turbulence: the total energy, the cross-helicity and the magnetic helicity.

%%%%%%%%%%%%%%%%%%%%%%%%%%%%%%%%%%%%%%%%%%%%%%%%%%%%%%%%%%%%%%%%%%%%%%%%%%%%%%%%
%%%%%%%%%%%%%%%%%%%%%%%%%%%%%%%%%%%%%%%%%%%%%%%%%%%%%%%%%%%%%%%%%%%%%%%%%%%%%%%%
\section{\turbo solver} \label{sec2}

The \turbo code\footnote{The code can be downloaded freely from: {\url{http://aqua.ulb.ac.be/turbo}}} is designed to solve the equations for an incompressible fluid in a three dimensional slab geometry with periodic boundary conditions in the three directions. The real space representation of the incompressible MHD equations under the influence of an external force and in the presence of a constant magnetic field $\mathbf{B}_0$ are written as,
\begin{align}
&\frac{\partial \uu}{\partial t} = - \uu \cdot \nabla \uu +\bb \cdot \nabla \bb +\mathbf{B}_0 \cdot \nabla \bb +\nu \nabla^2 \uu + \ff^u  - \nabla p + \ff^C\;, \label{vel}\\
&\frac{\partial \bb}{\partial t} = - \uu \cdot \nabla \bb +\bb \cdot \nabla \uu +\mathbf{B}_0 \cdot \nabla \uu +\eta \nabla^2 \bb +\ff^b \;, \label{mag} \\
&\nabla \cdot \uu=0\; ,\ \nabla \cdot \bb=0 \;, \label{divfree}
\end{align}
where $\uu=\uu(\xx,t)$ is the fluid velocity field, $\bb=\bb(\xx,t)$ is the magnetic field expressed in Alfv\`{e}n units and $p=p(\xx,t)$ is the total, hydrodynamic + magnetic, pressure field divided by the constant mass density. Due to the incompressibility condition, the pressure $p$ is not an independent variable and can be formally eliminated by solving the Poisson equation,
\begin{align}
\nabla^2 p=-\nabla \uu : \nabla \uu+\nabla \bb : \nabla \bb\; . 
\end{align}
The kinematic fluid viscosity is $\nu$ and the magnetic diffusivity is $\eta$. The divergence free external force fields $\ff^u=\ff^u(\xx,t)$ and $\ff^b=\ff^b(\xx,t)$ act on the velocity and magnetic fields, respectively. The two forces are part of a forcing mechanism that imposes the injection rates of the MHD ideal invariant quantities. A kinetic only forcing method ($\ff^b \equiv 0$), used previously in the literature for similar studies \cite{ Carati:2006p632, Teaca:2009p628}, can also be employed. In the velocity evolution equation, $\ff^u$ can be considered as divergence free since the pressure will enforce the incompressibility of the velocity field by eliminating any $\nabla\cdot\ff^u$ contribution of the force. On the other hand, $\ff^b$ must always be divergence free as a consistency condition for the magnetic field. 
If desired, a Coriolis force $\ff^C=\mathbf{\Omega} \times \uu$ acting on the flow can also be considered. The force appears as result of a reference system rotation, with the angular velocity $\mathbf{\Omega}$. In that case, the centrifugal acceleration that depends explicitly on the distance to the rotation axis can be lumped into the pressure term due to the incompressibility condition and the use of periodic boundary conditions is still appropriate. 

In addition to the MHD equation, the \turbo code can also solve the evolution equations for a set passive scalars fields:
\begin{align}
&\frac{\partial c_{\alpha }}{\partial t} = - \uu \cdot \nabla c_{\alpha } +\kappa _{\alpha}\nabla ^{2}c_{\alpha }+\sigma _{\alpha }(\{c_{\beta }\})  \label{scalareq}
\end{align}
where $c_{\alpha }=c_{\alpha }(\xx,t)$ are passive scalar(s), each of which is characterized by a diffusion coefficient $\kappa_\alpha$. There is the possibility to include source or sink terms or even chemistry terms in the scalar equations through the function $\sigma_\alpha$. 

In the \turbo code, the space discretization is based on a Fourier representation of the quantities of interest. For a given quantity, the physical $Q$ and the spectral $\hat{Q}$ representations are related using the direct and the inverse discrete Fourier transforms\footnote{Numerically, a Fast Fourier Transform algorithm is employed through the use of the FFTW  libraries, \cite{fftw}, which can be downloaded at: \url{http://www.fftw.org/}}: 
\begin{align}
\hat{Q}\mathbf{(k}) &=\frac{1}{N^{3}}\sum\limits_{\mathbf{x}}Q(\mathbf{x})e^{-i \kk \cdot \xx}\,, \\
Q(\mathbf{x}) &=\sum\limits_{\mathbf{k}}\hat{Q}(\mathbf{k})e^{i\kk \cdot \xx}\,.
\end{align}
where $N$ is the total number of modes in a given direction. In practice, the code allows the use of different numbers of modes in the three directions $N_x$, $N_y$ and $N_z$ and different box sizes $L_x$, $L_y$ and  $L_z$. Knowing the number of modes and the box size for each direction allows us to define the wavenumber space. Assuming a cubic box for simplicity of notations, the wavenumbers are defined as:
\begin{align}
k_n=\frac{2\pi}{L}n\; ,
\end{align}
where $n\in[-N/2+1,\ N/2]$. From the above definition, we see that the smallest nonzero wavenumber is $k_0=2\pi/L$, which for the typical box length choice $L=2\pi$, becomes unity. The largest wavenumber accounted in a simulation ($k_{\max}=\pi N/L$) and the spacing between two neighbour grid points ($\Delta=L/N$) are related as $\Delta=\pi/k_{\max}$. 

The main advantages of spectral methods are definitely the accuracy and the simplicity of the representation of the differentiation operator $\nabla$, which reduces to a multiplication in spectral space, $\nabla\, \uu(\xx) \rightarrow i\,\kk\, \hat \uu(\kk)$. The weak point of this method is the mandatory use of periodic boundary conditions, a choice considered as hard coded in the \turbo\ solver. Spectral methods are thus not adequate for exploring very complex geometries but are extremely useful for investigating the fundamental properties of turbulence. 

The time evolution is based on a modified Williamson, four-step, third-order low storage Runge-Kutta method~\cite{Williamson:1980p1043}. Since the equations are discretized in space, it is desirable that the transfer of information be limited to neighbor grid point in one time step $\Delta t$. This implies the CLF criterium which simply states that the time step $\Delta t$ has to be smaller than the time necessary for the wave with the largest propagation speed (determined on the velocity $\| \uu \|^{\max}$ or on the Alfv\'en velocity $\| \bb \|^{\max}$) to propagate between the smallest distance present between two grid points ($\Delta$). The linear terms are solved in an analytical manner for each mode $\kk$ by performing an appropriate change of variables $\hat \uu \rightarrow \hat \uu'(\hat \uu, \kk, \nu, \Omega)$\footnote{The operator of this tensorial transformation reduces to a diagonal operator for $\Omega=0$. } and $\hat \bb \rightarrow \hat \bb'(\hat \bb, \kk, \eta)$ before formally performing the Taylor series expansion. Although this will still give us a third-order accuracy in a global sense for the solution, the linear terms do not affect the value of the time step. 

Since the evolution equations are solved is spectral space, the value of the nonlinear terms for a mode $\kk$ has to be computed. Determining the nonlinear terms directly in Fourier space, although possible, is prohibitive from a computational cost point of view, as they are represented by convolutions of modes. Instead, inverse Fourier transforms are used for the velocity and the magnetic field and the nonlinear terms are computed in real space before being  Fourier transformed back to the spectral representation. In practice, the nonlinear terms appearing in the evolution equation for a velocity mode correspond to the divergence of a symmetric tensor $u_i u_j-b_i b_j$, while the nonlinear term appearing in the evolution equation for a magnetic field mode appears as the divergence of an anti-symmetric tensor $u_i b_j-b_i u_j$. This has an impact on the number of Fourier transforms that need to be perform: six for the symmetric tensor and only three for the anti-symmetric tensor, instead of nine transforms for each term. Furthermore, the trace of the tensor $u_i u_j-b_i b_j$ can be lumped into the pressure term, leading to another reduction of the computational efforts. Two orthogonal projections with respect to $\kk$ of the resulting nonlinear terms ensure that the solenoidal conditions are enforced for both the velocity and magnetic fields. 

Computing the nonlinear terms in real space reduces the numerical method type from a purely spectral method to a pseudo-spectral one. This approach of computing the nonlinear terms, although much faster then computing the convolution, adds another problem known as aliasing. The aliasing error originates in the fact that a plane wave with a wavenumber $k$ takes exactly the same values on the grid as a plane wave with a wavenumber $k+Nk_0$. When the nonlinearities are computed in real space, the aliasing error becomes a serious issue. For a one dimensional quantity $q$ represented by modes with $k=k_0\times[(-N/2+1),\ N/2]$, its square $q^2$ has modes that correspond to $k=k_0 \times[(-N+2), N]$. As a consequence, the mode of $q^2$ that corresponds to $k=k_0(-N+2)$ is undistinguishable from the mode $k=2k_0$. Two approaches, known as dealiasing methods, are considered to eliminate this difficulty~\cite{Patterson:1971p965}. 

The first method consists in assuming that only the modes of $q$ with $k=k_0 \times[(-M/2+1),\ M/2]$ are non zero and imposing that the other modes remain zero after each nonlinear computation. The {\em two-third method} consists simply taking $M=2/3\ N$ and keeping all the modes outside the range $k=k_0\times [(-N/3+1),\ N/3]$ to zero. This method represents the simplest way of fully removing the aliasing error.

The second method is based on the property that shifting the grid by a distance $d$ results in a modification for the modes by a phase $e^{i\, k\,d}$. For one-dimensional systems, the aliasing error can be removed exactly by computing the nonlinearity twice on two grids shifted respectively by $\Delta/2$ and $-\Delta/2$ and by summing the two computations. The contributions that do not lead to aliasing errors are unaffected by this procedure. In three dimensional systems, height evaluations with different shifts are needed to get an alias-free computation of the nonlinear terms. This is of course quite prohibitive and an approximation is made. The height shift computations of the nonlinear terms are done as part of the sub-step of a Runge-Kutta scheme. The {\em phase-shift dealiasing method} represents the reason behind the four step implementation for the third order Williamson Runge-Kutta time advancement method. This dealiasing method represents an approximative process of removing the aliasing errors.

The \turbo code is parallelized through the use of MPI. The numerical cube is split in the $y$-direction in spectral space, which limits the maximum number of processors used for one run to $N_y$. To improve the use of available parallel computing resources, an additional parallelization direction is made over the number of instances performed during each run, i.e. number of ``cubes'' solved at the same time that can exchange data among themselves. This last feature of the \turbo code allows the direct computation of ensemble average quantities.

%%%%%%%%%%%%%%%%%%%%%%%%%%%%%%%%%%%%%%%%%%%%%%%%%%%%%%%%%%%%%%%%%%%%%%%%%%%%%%%%
%%%%%%%%%%%%%%%%%%%%%%%%%%%%%%%%%%%%%%%%%%%%%%%%%%%%%%%%%%%%%%%%%%%%%%%%%%%%%%%%
\section{MHD ideal invariants}  \label{sec3}

In MHD turbulence, the three quadratic\footnote{Other invariants may exist, but only quadratic invariants are robust enough to survive truncation due to numerical discretization~\cite{Biskamp:2003}.} ideal invariants have an important role in the dynamics of turbulence and the resulting cascades that appear for fully developed turbulence regimes~\cite{Lithwick:2007p767, Beresnyak:2008p573}. For ideal MHD, that is an inviscid flow with zero magnetic diffusivity, total energy, cross-helicity and magnetic-helicity are conserved in absence of forcing. Magnetic-helicity $H^m=\langle \mathbf{a}\cdot \bb \rangle$ is a purely magnetic invariant, defined as the scalar product between the magnetic potential $\mathbf{a}$ and the magnetic field $\bb=\nabla \times \mathbf{a}$, where $\langle \dots \rangle$  denotes volume average. 
%
%%%%%%%
\begin{figure}
\centering
\includegraphics[width = 0.98\columnwidth]{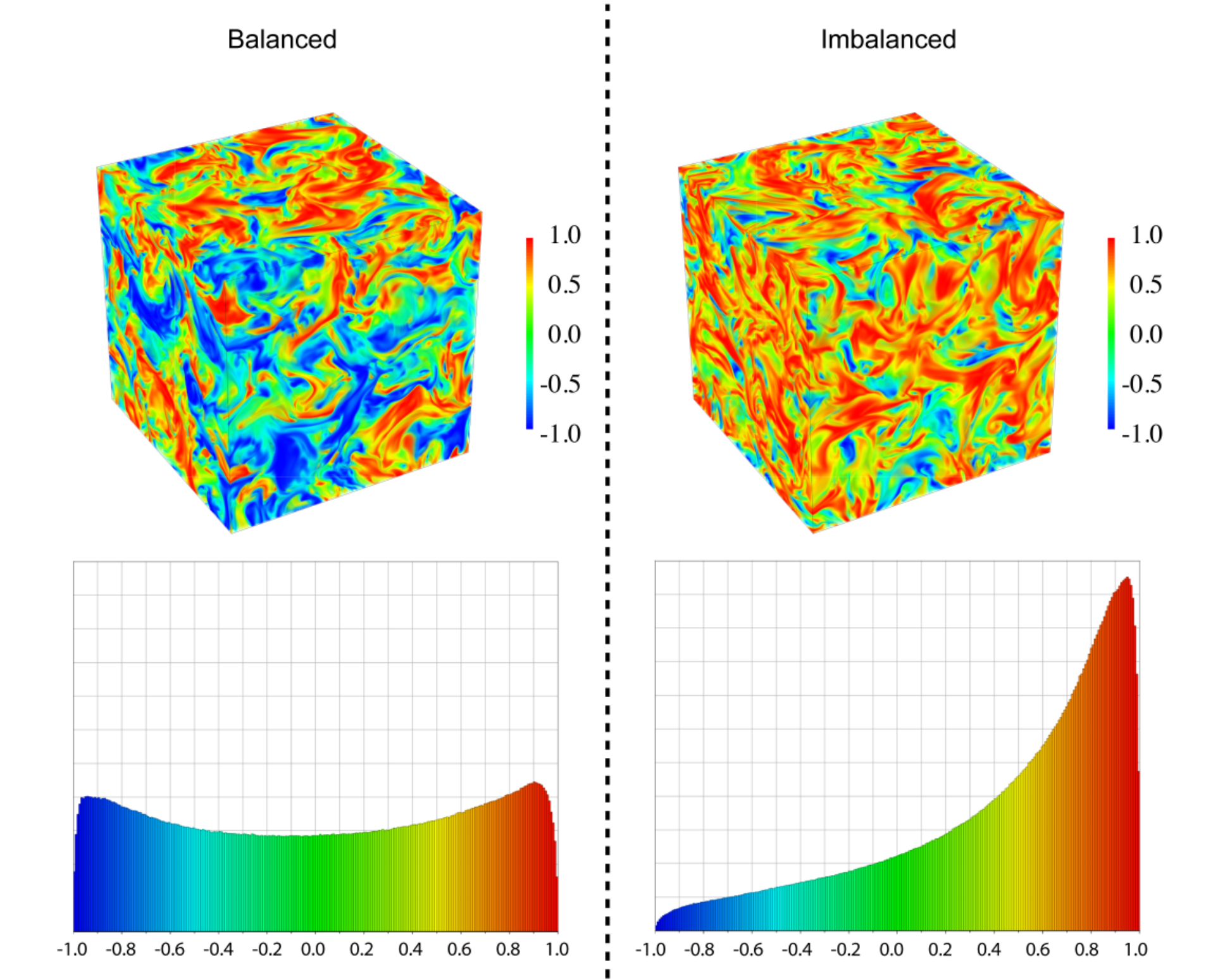}
\caption{Example of the real-space density of cross-helicity level for balanced and imbalanced MHD turbulence solved for $256$ modes in each direction. For the balanced case $\rho^c\approx0$ while for the imbalanced case $\rho^c\approx0.6$. The lower panels depict histograms of the cross-helicity level over the entire computational domain.     \label{fig01}}
\end{figure}
Since it possesses an inverse cascade (transfer form small to the large scales~\cite{Frisch:1975p841}), magnetic-helicity plays an important role in dynamo phenomena, see \cite{Alexakis:2006p298, Alexakis:2007p747}. The other two quadratic invariants, total energy $E=E^u+E^b$ and cross-helicity $H^c=\langle \uu \cdot \bb \rangle$ should be studied together as they affect each-other. Since kinetic $E^u=\langle \uu \cdot \uu \rangle/2$ and magnetic $E^b=\langle \bb \cdot \bb \rangle/2$ energies are not conserved individually, the use of Els\"asser variables $\zz^\pm=\uu \pm \bb$ might be more appropriate. In fact, in the Els\"asser representation, the cross-helicity and total energy information is contained in the definition of two ideal invariants $E^+=\langle \zz^+ \cdot \zz^+ \rangle/4$ and $E^-=\langle \zz^- \cdot \zz^- \rangle/4$, known as pseudo-energy. The $E^\pm$ ideal invariants are positively defined, which represents an advantage in spectral study. Since cross-helicity is related to the degree of alignement between the velocity and the magnetic field, the cross-helicity level defined as,
\begin{align}
&\rho^c = \frac{H^c}{E} \equiv \frac{(E^+-E^-)}{(E^++E^-)}\;, \label{cross}
\end{align}
provides global information regarding the alignment present in a system ($\rho^c\in[-1,1]$). Point-wise, the presence of a large cross-helicity level $\rho^c(\xx)=H^c(\xx)/E(\xx)$ gives rise to the phenomena of nonlinear depletion in the evolution equations. Nonlinear depletion weakens the ability of the nonlinear terms to mix the flow compared to hydrodynamic case, which results in a different turbulent mixing time for MHD turbulence~\cite{Zhou:2005p976}. Moreover, it was shown in \cite{Matthaeus:2008p53} that MHD turbulence is composed of zones of highly aligned ($\rho^c(\xx)\sim+1$) and highly anti-aligned structures ($\rho^c(\xx)\sim-1$) which are thought to develop naturally through the process of dynamical alignment \cite{Boldyrev:2009p820}. Therefore, a non-zero value for the global parameter $\rho^c$ denotes a preference in the generation of one type of aligned structures. This situation is known as imbalanced MHD turbulence, Figure~\ref{fig01}. It is interesting to note that the presence of cross-helicity does not necessarily affect the equipartition of energy between the $\uu$ and $\bb$ fields. However, it does change the pseudo-energy levels of $\zz^+$ and $\zz^-$ and therefore the strength of the co-propragating and contra-propagating Alfv\`{e}n waves that scatter on each-other. 

Although the ideal invariants are not conserved in the presence of viscosity and magnetic diffusivity, they are still redistributed between different scales without loss or gain through their nonlinear fluxes. For $X$, $Y$, $Z$ standing in for the fields, the flux through a spherical surface in the Fourier space defined by a radius $k_c$ is defined as: 
\begin{align}
\Pi^X_{Y,Z}(k_c)= \langle \mathbf{Z}(\kk) \cdot \nabla \mathbf{Y}(\kk|k<k_c) \cdot  \mathbf{X}(\kk|k>k_c) \rangle\;.
\end{align}
The average represents here the sum over all the Fourier modes. For a turbulent state, the value of the fluxes in the inertial-inductive range are expected to be constant. In the Els\"asser formalism, the values $\Pi^+_{+,-}$ and $\Pi^-_{-,+}$ reached in the inertial-inductive range, enter in the phenomenological definition of pseudo-energy scaling laws~\cite{Verma:2004p206}.

%%%%%%%%%%%%%%%%%%%%%%%%%%%%%%%%%%%%%%%%%%%%%%%%%%%%%%%%%%%%%%%%%%%%%%%%%%%%%%%%
%%%%%%%%%%%%%%%%%%%%%%%%%%%%%%%%%%%%%%%%%%%%%%%%%%%%%%%%%%%%%%%%%%%%%%%%%%%%%%%%
\section{The forcing mechanism} \label{sec4}

\begin{figure}
\centering
\includegraphics[width = 0.98\columnwidth]{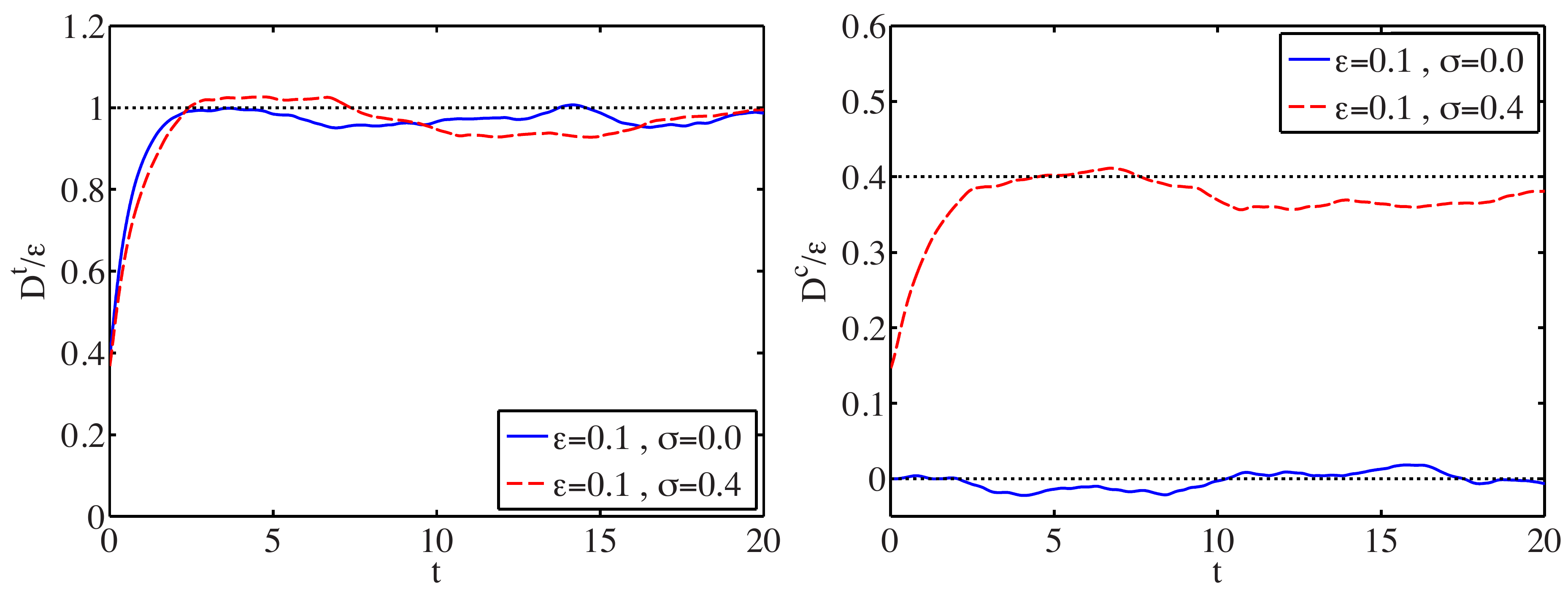}
\caption{The total energy dissipation ($D^t$) level (left) and and cross-helicity dissipation ($D^c$) (right) for two different $\sigma$ values. The plots are made for well resolved turbulence using a numerical resolution of $512$ in each direction. \label{fig02}}
\end{figure}

As it is known, a forcing mechanism can be used to reach different turbulent regimes. In this study, the injection rates of the nonlinear ideal invariants will be used as the control parameters. These injection rates can easily be computed from equations (\ref{vel}-\ref{mag}). For instance, the injection rates of total energy and of cross-helicity ($H^c=\langle \Re\{\hat\uu(\kk) \cdot \hat\bb(-\kk)\} \rangle$) are given by
\begin{align}
\left. \frac{\partial E}{\partial t} \right|_f     &=\langle \hat\ff^u(\kk) \cdot \hat\uu(-\kk) \rangle+\langle \hat\ff^b(\kk) \cdot \hat\bb(-\kk) \rangle = \varepsilon^u+\varepsilon^b=\varepsilon \;, \\
\left. \frac{\partial H^c}{\partial t} \right|_f &=\langle \hat\ff^u(\kk) \cdot \hat\bb(-\kk) \rangle+\langle \hat\ff^b(\kk) \cdot \hat\uu(-\kk) \rangle =  \sigma^u+ \sigma^b=\varepsilon \sigma\;, \label{crossinj}
\end{align}
where, thanks to the Parseval theorem, the volume average $\langle \dots \rangle$ can be identified as the average over the number of modes.
The parameters $\varepsilon^u$ and $ \varepsilon^b$ represent the power injected by $\hat\ff^u$ and $ \hat \ff^b$ respectively. Since the cross-helicity is bounded by the total energy, the sum of $\sigma^u$ and $ \sigma^b$, which individually denote the cross-helicity injected by $\hat\ff^u$ and $ \hat \ff^b$ respectively, has to fulfil  the condition: $-\varepsilon\le\sigma^u+\sigma^b\le+\varepsilon$. As such, the cross-helicity parameter $\sigma$ is bounded in the interval $[-1,1]$. Selecting the force control parameters in such a way to fix $\varepsilon$ and $\sigma$ will enforce the dissipation level for the energy and cross-helicity, once the stationary regime is reached as shown on Figure~\ref{fig02}. This behavior is true only for ideal invariant quantities. For example, selecting the injection level of kinetic helicity will not enforce the kinetic helicity dissipation for MHD turbulence as it would in a purely hydrodynamic flow. 

Numerically, we consider the forces $\hat{\ff}^u(\kk)$ and $\hat{\ff}^b(\kk)$ to be local quantities in Fourier space, which act in the same manner on all the modes $N_{f}$ within a wavenumber shell defined by the interval $s_{f}=[k_{\inf }, k_{\sup}]$. Since $\hat \uu$, $\hat \bb$ and $\hat \ff^{u,b}$ are divergence free, we use a helical decomposition~\cite{Waleffe:1998p544} for the definition of the force. The helical decomposition projects a vector $\hat{\mathbf{a}}(\kk)$ on a complex basis $\mathbf{h}_\pm$:
\begin{align}
\hat{\mathbf{a}}(\kk) &=\hat{a}_+(\kk)\mathbf{h}_+ + \hat{a}_-(\kk) \mathbf{h}_-\;,
\end{align}
where $\mathbf{h}_\pm=\mathbf{e}_1 \times \mathbf{e}_2  \pm i \mathbf{e}_1$ and $\mathbf{e}_1=(\mathbf{\lambda} \times \mathbf{e}_2)/ \| \mathbf{\lambda} \times \mathbf{e}_2 \|$, $\mathbf{e}_2= \kk/ k$. The wave-vector $\mathbf{\lambda}$ is taken to be arbitrary and non-parallel to $\kk$. We warn the reader that in this section, the $\pm$ lower indices refer to the helical basis $h_\pm$, the vector projections on this basis and their contributions to the different scalar quantities of interest, while the upper indices denote, as usual, quantities in the Els\"asser representation. The helical decomposition is very useful in ensuring zero divergence and in computing the curl operator. Indeed, the vectors $\mathbf{h}_\pm$ are eigenmodes of the curl operator, $i\kk \times \mathbf{h}_\pm=\pm k \mathbf{h}_\pm$. As such the vorticity is now defined as $\hat \omega_\pm(\kk)=\pm k \hat u_\pm(\kk)$ and the electric current has the form $\hat j_\pm(\kk)=\pm k \hat b_\pm(\kk)$.

We choose the projection of the forces on $\mathbf{h}_\pm$, to have the form:
\begin{align}
\hat{\ff}^u_\pm(\kk) &=\alpha^u_\pm(\kk) \hat{u}_\pm(\kk) + \beta^u_\pm(\kk) \hat{b}_\pm(\kk) \;,\\
\hat{\ff}^b_\pm(\kk) &=\alpha^b_\pm(\kk) \hat{u}_\pm(\kk) + \beta^b_\pm(\kk) \hat{b}_\pm(\kk) \;,
\end{align}
if $ |\kk| \in s_{f}$ and zero otherwise. Since the $\alpha$'s and $\beta$'s parameters that need to be determined are considered to be real, the forcing method presented here does not influence the phases of the fields, which ensures that no change is made in the type of turbulent structures present. Another way of injecting cross-helicity into the system would be achieved by imposing the alignment of $\uu$ and $\bb$ in the real space, which in turn would modify the phases of the fields and potentially the turbulence behavior.

%%%%%%%

For a mode $\kk$, the kinetic energy ($E^u(\kk)=\frac{1}{2}\hat{\uu}(\kk) \cdot \hat{\uu}(-\kk)$), magnetic energy ($E^b(\kk)=\frac{1}{2}\hat{\bb}(\kk) \cdot \hat{\bb}(-\kk)$), cross-helicity ($H^c(\kk)=\Re \{ \hat{\uu}(\kk) \cdot \hat{\bb}(-\kk) \}$), kinetic-helicity ($H^k(\kk)=\Re \left\{ \hat{\uu}(\kk) \cdot \hat{\mathbf{\omega}}(-\kk) \right\}$) and the magnetic-helicity ($H^m(\kk)=\Re \{ \hat{\bb}(\kk) \cdot \hat{\mathbf{a}}(\kk)^{\ast} \}=\displaystyle{\frac{1}{k^2}\Re \{ \hat{\bb}(\kk) \cdot \hat{\mathbf{j}}(-\kk) \}}$) can be easily expressed in terms of the helical decomposition. Knowing that for the helical decomposition, the energy injected per unit of time in the velocity equation and the magnetic equation are $\varepsilon^u=\varepsilon^u_+ +\varepsilon^u_-$ and $\varepsilon^b=\varepsilon^b_+ +\varepsilon^b_-$ and that the cross-helicity injected per unit of time in the velocity equation and the magnetic equation are $\sigma^u= \sigma^u_+ +\sigma^u_-$ and $\sigma^b=\sigma^b_+ +\sigma^b_-$, respectively, we can write the injection rates for the three ideal invariants and kinetic helicity per mode $\kk$ in the forcing range as: 
\begin{align}
\left. \frac{\partial E(\kk) }{\partial t}\right|_f     & = \frac{1}{N_f}[(\varepsilon^u_+ + \varepsilon^u_-) + (\varepsilon^b_+ +  \varepsilon^b_-)]=\frac{1}{N_f}\varepsilon \;, \\
\left. \frac{\partial H^c(\kk) }{\partial t} \right|_f &= \frac{1}{N_f}[(\sigma^u_+ + \sigma^u_-) + (\sigma^b_+ +  \sigma^b_-)]=\frac{1}{N_f}\varepsilon \sigma\;, \\
\left. \frac{\partial H^m(\kk) }{\partial t} \right|_f &= \frac{1}{N_f} \frac{1}{k}(\varepsilon^b_+-\varepsilon^b_-)\;,\\
\left. \frac{\partial H^k(\kk) }{\partial t} \right|_f &=\frac{1}{N_f} k(\varepsilon^u_+ -\varepsilon^u_-) \;.
\end{align}

To simplify the numerical implementation of the force, we make the following assumption: we consider $\sigma^u_\pm=\sigma \varepsilon^u_\pm$ and $\sigma^b_\pm=\sigma \varepsilon^b_\pm$. In practice, these equalities impose that both forcing $\hat\ff^u$ and $ \hat \ff^b$ are responsible for the same amount of cross helicity injection in the system. Moreover, they also impose that cross helicity is injected at the same rate in both the $\mathbf{h}_+$ and $\mathbf{h}_-$ components of the velocity and magnetic fields. These assumptions, which could be reconsidered in the future, allow us to fix the eight real parameters $\alpha^u_\pm(\kk)$, $\beta^u_\pm(\kk)$, $\alpha^b_\pm(\kk)$ and $\beta^b_\pm(\kk)$ by giving only five control parameters, namely the energy injection rates $\varepsilon^u_\pm$, $\varepsilon^b_\pm$  and the cross-helicity parameter $\sigma$,
\begin{align}
\alpha^u_\pm(\kk) &=\frac{\varepsilon^u_\pm}{N_f}\frac{\sigma H^c_\pm(\kk)^2-2 E_\pm^b(\kk)}{H^c_\pm(\kk)^2-4E_\pm^u(\kk) E_\pm^b(\kk)} \;,\\
\beta^u_\pm(\kk)   &=\frac{\varepsilon^u_\pm}{N_f}\frac{\sigma H^c_\pm(\kk)^2-2 E_\pm^u(\kk)}{H^c_\pm(\kk)^2-4E_\pm^u(\kk) E_\pm^b(\kk)} \;,\\
\alpha^b_\pm(\kk) &=\frac{\varepsilon^b_\pm}{N_f}\frac{\sigma H^c_\pm(\kk)^2-2 E_\pm^b(\kk)}{H^c_\pm(\kk)^2-4E_\pm^u(\kk) E_\pm^b(\kk)} \;,\\
\beta^b_\pm(\kk)   &=\frac{\varepsilon^b_\pm}{N_f}\frac{\sigma H^c_\pm(\kk)^2-2 E_\pm^u(\kk)}{H^c_\pm(\kk)^2-4E_\pm^u(\kk) E_\pm^b(\kk)} \;, 
\end{align}
where the respective injection rates are assumed to be the same for all the $N_f$ forced modes. Selecting a large number of forced modes ensures that no anisotropy effect is induced by the forcing mechanism. Because of the Cauchy-Schwarz inequality, we have the condition $H^c_\pm(\kk)^2 \leq 4 E^u_\pm(\kk) E^b_\pm(\kk)$. We see that for the equality case, we develop a pole for the parameters $\alpha^u_\pm(\kk)$, $\beta^u_\pm(\kk)$,$\alpha^b_\pm(\kk)$ and $\beta^b_\pm(\kk)$. In an effort to advert this, we consider the condition $\sigma \varepsilon_\pm \le\sqrt{4\varepsilon^u_\pm \varepsilon^b_\pm}$ on the control parameters, where $\varepsilon_\pm=\varepsilon^u_\pm+\varepsilon^b_\pm$.

Since the injection rates per mode for the kinetic-helicity and magnetic helicity depend on $k$, the global kinetic-helicity injection rate $h$  and the global magnetic-helicity injection rate $\chi$ are found as, 
\begin{align}
\left. \frac{\partial H^k}{\partial t} \right|_f &=(\varepsilon^u_+ -\varepsilon^u_-)\sum_{s_f}k= h \;,\\
\left. \frac{\partial H^m}{\partial t} \right|_f &=(\varepsilon^b_+-\varepsilon^b_-)\sum_{s_f}\frac{1}{k}=\chi \;.
\end{align}

From the above expression of the forces, we find the forces that act in the Els\"asser form of the MHD equations as $\hat \ff^\pm=\hat \ff^u \pm \hat \ff^b$. If we chose a purely mechanical forcing of the turbulence ($\hat \ff^b=0$) for this forcing method, we obtain the relation $\hat \ff^+=\hat \ff^-=\hat \ff^u$. For this case $\varepsilon^b=0$, which numerically is found to be unstable unless no cross-helicity is injected, $\sigma=0$.
 
As a note, we see that using the helical decomposition we can redefine the hydrodynamical force we used in previous studies \cite{Teaca:2009p628, Carati:2006p632}  ($\ff^b \equiv 0$). From conditions imposed on the energy and kinetic helicity injection levels, we take the projections on $h_\pm$ of the force to be: 
\begin{align}
\hat{\ff}_\pm(\kk)  &=\frac{\varepsilon_\pm}{N_f}\frac{\hat u_\pm(\kk)}{E_\pm^u(\kk)}\;.
\end{align}
The condition for the kinetic helicity injection rate is automatically fulfilled for any selection of $\varepsilon_\pm$.

%%%%%%%
\begin{figure}
\centering
\includegraphics[width = 0.98\columnwidth]{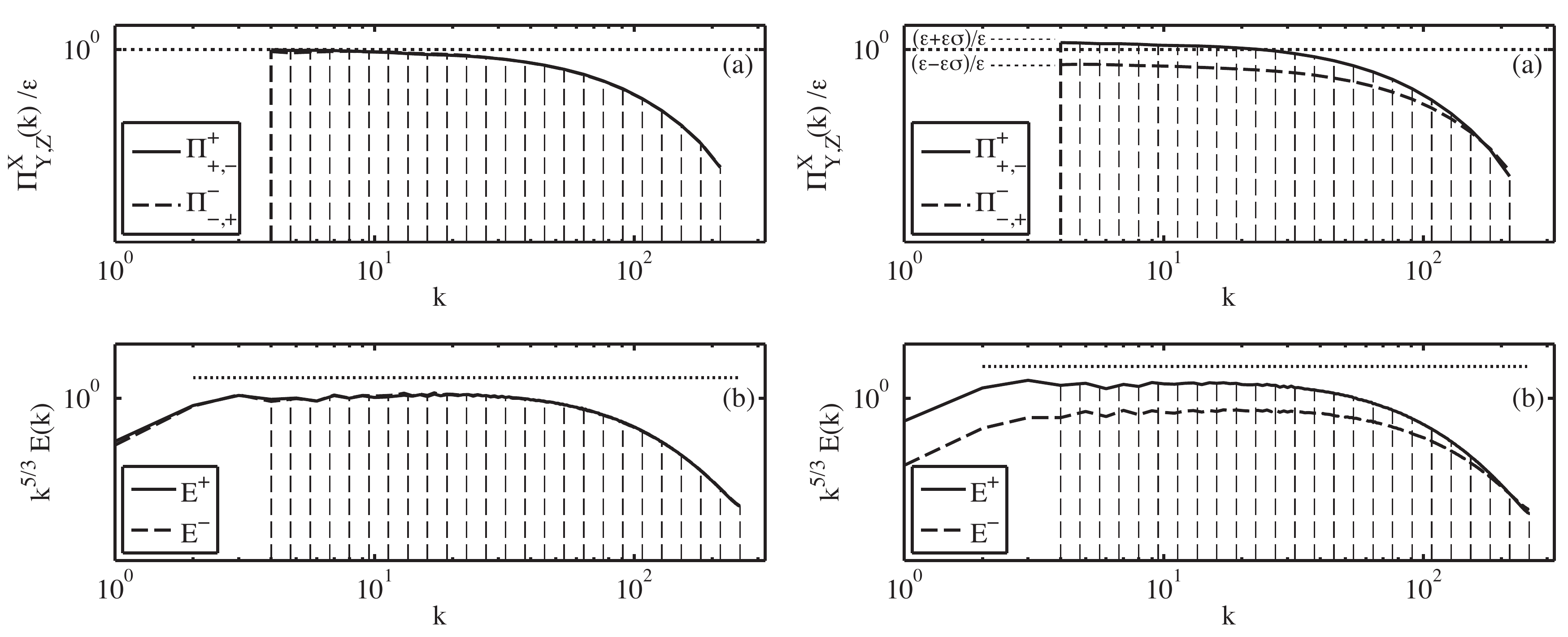}
\caption{Spectra (b) and fluxes (a) at the last time point computed in Figure~\ref{fig02}, for the Els\"asser pseudo-energy. Left panels depict balanced turbulence ($\sigma=0$; $\rho^c=0$) while right panel depict imbalanced turbulence ($\sigma=0.4$; $\rho^c=0.6$). \label{fig03} }
\end{figure}

%%%%%%%%%%%%%%%%%%%%%%%%%%%%%%%%%%%%%%%%%%%%%%%%%%%%%%%%%%%%%%%%%%%%%%%%%%%%%%%%
%%%%%%%%%%%%%%%%%%%%%%%%%%%%%%%%%%%%%%%%%%%%%%%%%%%%%%%%%%%%%%%%%%%%%%%%%%%%%%%%
\section{Conclusions and discusion}

A forcing mechanism which controls the injection level of the three ideal invariants of MHD turbulence: the total energy, the cross-helicity and the magnetic helicity has been developed and results obtained from the implementation of this force into the \turbo solver have been presented. This type of force represents a useful tool for the spectral study of turbulence, since it allows to control the level to which all three ideal quadratic MHD invariant fluxes relax to. 

As seen from Figure~\ref{fig03}, the presence of a non-zero cross-helicity injection level causes a change in the two pseudo-energy dissipation rates, which in turn causes a separation of the respective fluxes levels once a statistical stationary regime is reached. Looking at the spectra of $E^+$ and $E^-$, we observe a $5/3$ scaling. For imbalanced MHD turbulence, we also observe a difference in the pseudo-energy spectra levels. However, the scaling exponent tends to remain $5/3$. 

By controlling the cross-helicity injection level, this type of force should provide a help in the study of solar wind turbulence, which represents a well known case of imbalanced MHD turbulence. Also, controlling the magnetic helicity injection level should be of great help in the field of galactic and solar dynamo physics.

%%%%%%%%%%%%%%%%%%%%%%%%%%%%%%%%%%%%%%%%%%%%%%%%%%%%%%%%%%%%%%%%%%%%%%%%%%%%%%%%
%%%%%%%%%%%%%%%%%%%%%%%%%%%%%%%%%%%%%%%%%%%%%%%%%%%%%%%%%%%%%%%%%%%%%%%%%%%%%%%%
\acknowledgments{
B. Teaca would like to acknowledge Stefaan Poedts and Grigol Gogoberidze for discussions which lead to the development of this force, K.U.Leuven for the use of VIC3 cluster computational resources and Benjamin Cassart for the help provided in maintaining the \turbo code.}

%%%%%%%%%%%%%%%%%%%%%%%%%%%%%%%%%%%%%%%%%%%%%%%%%%%%%%%%%%%%%%%%%%%%%%%%%%%%%%%%
%%%%%%%%%%%%%%%%%%%%%%%%%%%%%%%%%%%%%%%%%%%%%%%%%%%%%%%%%%%%%%%%%%%%%%%%%%%%%%%%

%\bibliographystyle{Thesis_Reference}
%\bibliography{Thesis_Reference}

%%%%%%%%%%%%%%%%%%%%%%%%%%%%%%%%%%%%%%%%%%%%%%%%%%%%%%%%%%%%%%%%%%%%%%%%%%%%%%%%
%%%%%%%%%%%%%%%%%%%%%%%%%%%%%%%%%%%%%%%%%%%%%%%%%%%%%%%%%%%%%%%%%%%%%%%%%%%%%%%%
\end{document}